\documentclass[aps,pre, twocolumn,showpacs,superscriptaddress]{revtex4}

\usepackage{graphicx}
\usepackage{SIunits}

\newcommand{\imag}{\mathrm{i}}

\begin{document}

\title{Dynamical instabilities of dissipative solitons in nonlinear optical cavities with nonlocal materials}

\author{Lendert \surname{Gelens}}
\email[Electronic Address:]{lendert.gelens@vub.ac.be}
\affiliation{Department of Applied Physics and Photonics,
             Vrije Universiteit Brussel,
             Pleinlaan 2,
             B-1050 Brussel,
             Belgium}
             
\author{Dami\`a \surname{Gomila}}
\email[Electronic Address:]{damia@ifisc.uib.es}
\affiliation{IFISC, Instituto de F\'{\i}sica Interdisciplinar y Sistemas 
Complejos (CSIC-UIB), Campus Universitat Illes Balears, E-07122 Palma de
Mallorca,  Spain}
             
\author{Guy \surname{Van der Sande}}
\affiliation{Department of Applied Physics and Photonics,
             Vrije Universiteit Brussel,
             Pleinlaan 2,
             B-1050 Brussel,
             Belgium}
\affiliation{IFISC, Instituto de F\'{\i}sica Interdisciplinar y Sistemas 
Complejos (CSIC-UIB), Campus Universitat Illes Balears, E-07122 Palma de
Mallorca,  Spain}
             
\author{Jan Danckaert}
\affiliation{Department of Applied Physics and Photonics,
             Vrije Universiteit Brussel,
             Pleinlaan 2,
             B-1050 Brussel,
             Belgium}
             
\author{Pere \surname{Colet}}
\affiliation{IFISC, Instituto de F\'{\i}sica Interdisciplinar y Sistemas 
Complejos (CSIC-UIB), Campus Universitat Illes Balears, E-07122 Palma de
Mallorca,  Spain}
             
\author{Manuel A. \surname{Mat\'ias}}
\affiliation{IFISC, Instituto de F\'{\i}sica Interdisciplinar y Sistemas 
Complejos (CSIC-UIB), Campus Universitat Illes Balears, E-07122 Palma de
Mallorca,  Spain}

\date{\today}

\begin{abstract} 

In this work we characterize the dynamical instabilities of localized structures exhibited by a recently introduced [Gelens {\it et al.}, Phys. Rev. A {\bf
75}, 063812 (2007)] generalization of the Lugiato-Lefever model to include a weakly
nonlocal response of an intracavity metamaterial. A rich scenario, in which the
localized structures exhibit different types of oscillatory  instabilities, tristability, and excitability, including a regime of conditional excitability in which the system is bistable, is presented and discussed. Finally, it is shown that the scenario is organized
by a pair of Takens-Bogdanov codimension-$2$ points.

\end{abstract}

\maketitle

\section{Introduction} 
Dissipative solitons (DS) are spatially localized structures that appear in 
certain nonlinear dissipative media \cite{Thual88,Akhmediev05}. They have been found in systems such as chemical reactions
\cite{Pearson-1993,Lee-1993}, vegetation models \cite{Lejeune-2006}, gas discharge
systems \cite{Muller-1999}, fluids \cite{Thual-1990} and optics
\cite{Scroggie-1994,Tlidi-1994,Taranenko-1997,2002Natur.419..699B,Bortolozzo-2006,
Trillo-2001,Rosanov-2002,Mandel-2004}. While in many instances these localized 
structures are stable, there are situations in which they develop different kinds of 
instabilities. Some instabilities lead to the formation of an extended pattern
and therefore the localized character of the DS is destroyed. More interesting
are the instabilities that, while preserving its localized character, induce the
DS to start moving, breathing or oscillating 
\cite{Firth96,Longhi97,2002JOSAB..19..747F,Vanag04}.
Since DS can be considered as an entity on their own, these instabilities may lead to dynamical regimes that appear not to be present in the dynamical behavior of the
extended system. In this context it has recently been reported that DS arising 
in a prototype model for optical cavities filled with a nonlinear Kerr media 
may show excitable behavior, while locally the system is not
excitable \cite{2005PhRvL..94f3905G, Damia_PRE}. Thus, excitability can be an 
emergent property arising from the spatial dependence, which allows for the 
formation of localized structures. In that situation excitability is mediated
by a saddle-loop bifurcation and the whole scenario is organized by a
Takens-Bogdanov (TB) codimension-2 point. In parameter space the TB point is
located in the asymptotic limit in which the model becomes equivalent to the
Nonlinear Schr\"odinger Equation (NLSE).

Since this excitability scenario is an emergent property of the spatial
dependence of the system, it is particularly important to characterize how this
scenario may change when the nature of spatial coupling is varied. In the
Lugiato-Lefever model \cite{LL-1987} considered in Refs. 
\cite{2005PhRvL..94f3905G, Damia_PRE} the spatial coupling arises from optical
diffraction in the paraxial approximation and is therefore accounted for by a
Laplacian term. Here, we consider an extension of the model, including a 
mildly nonlocal term which extends the range of spatial interaction 
\cite{Gelens_PRA_2007}.  This extension of the original
Kerr model is suggested by the recent availability of metamaterials, allowing to design an optical Kerr cavity where layers of right- and left-materials 
are alternated. This provides the possibility to strongly decrease the 
diffraction strength in the resonator, such that higher order spatial effects 
(e.g. nonlocal effects) start to dominate the dynamical behavior of the DS. In 
this work, it will be shown that the additional spatial interaction term is 
able to shift the bifurcation lines such that now two Takens-Bogdanov points 
move from infinity to finite parameter values, acting as organizing centers of a richer 
dynamical behavior.

\section{Model}
We consider an optical cavity with a Kerr-type nonlinearity, driven by a
homogeneous, coherent optical light beam. This system was first introduced
by Lugiato and Lefever to study pattern formation in a driven nonlinear, passive
optical resonator \cite{LL-1987}. In this work, we will study a more general
equation, which includes a bilaplacian term:
\begin{equation}
\frac{\partial E}{\partial{t}}  = -(1+\imag\theta )E+E_\mathrm{in}
+\imag\,\vert{E}\vert^2\,E +\imag\alpha\nabla_\perp^2 E 
+ \imag\beta\nabla_\perp^4 E.
\label{Eq:LL_beta}
\end{equation}
This model equation has been obtained in Ref.\ \cite{Gelens_PRA_2007,
2006PhRvA..74c3822K,Tassin07} to describe the temporal evolution of the slowly varying
envelope of the electric field $E(\vec{x}, t)$ in a double-layered optical
cavity. One layer of the cavity consists of a conventional right-handed
material, while the other layer is an optical left-handed metamaterial. $\vec{x} = (x,y)$ represents the plane transverse to the propagation
direction. Eq.\ (\ref{Eq:LL_beta}) has been obtained under the same approximations under which
the Lugiato-Lefever equation is valid, i.e., the slowly varying envelope
approximation, weak nonlinearity, the paraxial limit and a nearly resonant
cavity. The bilaplacian term appears when taking into account a linear and weakly
nonlocal response of the left-handed metamaterial. Furthermore, it has been
shown that in this double-layered cavity the ratio between $\alpha$ and $\beta$
can be drastically altered by changing the relative lengths of both material
layers. Recent advances in the field of metamaterials \cite{smith2004,
Shalaev-2007}, potentially allow to find DS modeled by Eq.\
(\ref{Eq:LL_beta}) in a wide area of the parameter space. \\

The first term at the right-hand side of Eq.\ (\ref{Eq:LL_beta}) models the
cavity losses, $\theta$ represents the cavity detuning with respect to the
driving field $E_{\text{in}}$, $\nabla_\perp^2 = \partial^2 / \partial x^2 + \partial^2
/ \partial y^2$ is the transverse Laplacian term due to diffraction, and
$\nabla_\perp^4$ the transverse bilaplacian term modeling the linear, weakly
nonlocal response. The cubic self-focusing nonlinearity is given by
$\imag\,\vert{E}\vert^2\,E $. Eq.\ (\ref{Eq:LL_beta}) has a homogeneous steady
state solution $E_s = E_0 / (1+ \imag(\theta - I_s))$, where $I_s =
\vert{E_s}\vert^2$ \cite{LL-1987, Gelens_PRA_2007}.
The homogeneous solution becomes modulationally unstable at  $I_s = 1$, which is also the case for the regular
Lugiato-Lefever equation \cite{LL-1987}, but in Eq.\ (\ref{Eq:LL_beta}) the instability arises
with two characteristic wavenumbers \cite{Gelens_PRA_2007}. From this modulational instability, a
subcritical branch of DS appears. In the remainder of this manuscript, we
use the background intensity $I_s$, the detuning $\theta$ and the coefficient of
the bilaplacian term $\beta$, that is a measure of the  strength of the
nonlocality, as our control parameters. Without loss of generality, we take $\alpha = 1$. 

\section{Dynamical behavior}
 
\begin{figure}
\includegraphics[clip]{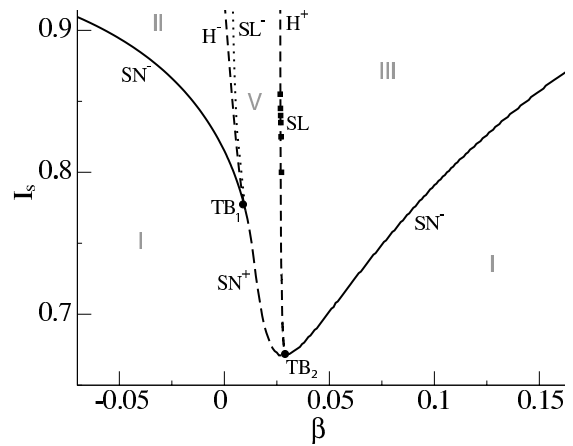}
\caption{Phase diagram of DS in the nonlocal Kerr cavity for $\theta = 1.23$. DS do not exist in Region I, below the saddle node line (SN).
DS are stable between the saddle-node bifurcation (solid line - SN) and the 
Hopf bifurcation (dashed line - H), namely in Regions II, III, VI and VIII
(see also Fig.~\ref{Fig::bif_beta_Is_ZoomTB1}).
In Regions IV and VII delimited by a Hopf and a saddle-loop bifurcation
(dotted line - SL) there are stable oscillatory DS while the stationary DS are 
unstable. In Region V the static DS is unstable and the phase space generated
after the destruction of the limit cycle at the SL induces a regime of 
excitable DS. Region VI corresponds to a regime of conditional excitability
with both stable and excitable DS. Finally in Region VIII one encounters
tristability: a stationary and an oscillatory DS coexist with the homogeneous
solution.
Where the saddle-node bifurcation line and the Hopf bifurcation line meet is a 
Takens-Bogdanov (TB) bifurcation.}
\label{Fig::bif_beta_Is}
\end{figure}

\subsection{Preliminary remarks}\label{Sect::PrelRemarks}

\begin{figure}
\includegraphics[clip]{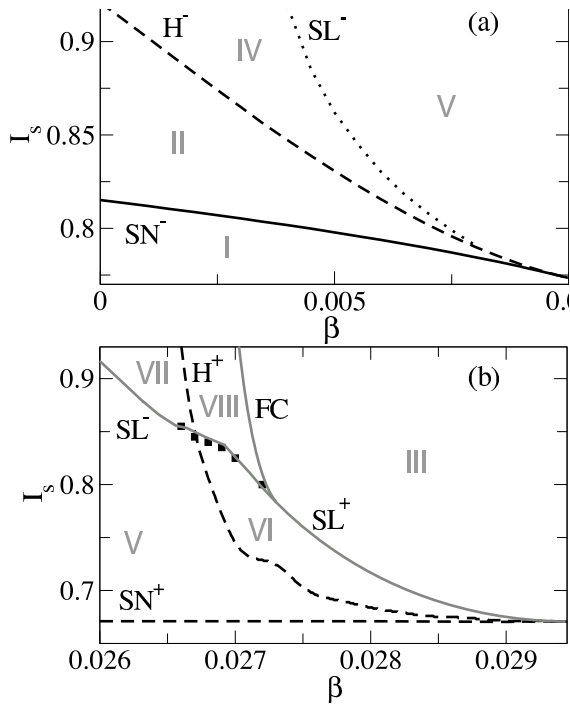}
\caption{Zoom of Fig.\ \ref{Fig::bif_beta_Is} near TB$_1$ (a) and TB$_2$ (b). 
Lines have been determined using the method explained in Section\ \ref{Sect::PrelRemarks} with the exception of the SL line in panel (b).
In that case the filled squares display the location of the SL 
obtained from numerical integration of Eq.\ (\ref{Eq:LL_beta}), while the grey line through these 
points is only to guide the eye. $\theta$ = 1.23.} 
\label{Fig::bif_beta_Is_ZoomTB1}
\end{figure}

In this section, we show the different possibilities of dynamical behavior of DS in parameter
space (see Fig.\ \ref{Fig::bif_beta_Is}). Since DS are radially symmetric they 
correspond to stationary solutions of the radial form of Eq. (\ref{Eq:LL_beta}) 
with boundary conditions $\partial_r E(r=0)=0$ and $\partial_r E(r \rightarrow \infty)=0$. 
This equation can be solved numerically using a Newton method \cite{McSloy, Damia_PRE, Gelens_PRA_2007}. This approach is very accurate and automatically 
generates the Jacobian operator, whose eigenvalues determine the stability of 
the solutions. Note that this method finds both stable and unstable stationary solutions. DS can undergo two kinds of instabilities, radial 
instabilities which preserve the localized character of the structure and
azimuthal instabilities which lead to the formation of extended patterns. 
The last ones appear only for large values of the background intensity 
($I_s$ close to 1). We focus here on the radial instabilities, so phase diagrams
(Figs.\ \ref{Fig::bif_beta_Is} and \ref{Fig::bif_beta_Is_ZoomTB1}) are plotted 
only up to $I_s=0.93$, before the azimuthal instabilities take place.

Since the system has three parameters ($I_s$, $\theta$, $\beta$), for the sake
of clarity we fix the detuning at $\theta=1.23$ in this section and 
analyze a slice of the whole parameter space. 
Fig.~\ref{Fig::bif_beta_Is} shows the
region of the parameter plane $(\beta, I_s)$ that contains the most relevant
regimes of dynamical behavior of the system. 
The line that dominates this parameter plane
has the shape of a deformed parabola, and is a line of saddle-node (SN) bifurcations in which two DS are created.
Below this line one has Region I where no DS exist. We recall 
that in all the parameter range covered
by Fig.~\ref{Fig::bif_beta_Is} the spatially homogeneous solution is always
stable. The different regimes above this line are organized by 
two codimension-$2$ Takens-Bogdanov points, 
TB$_1$ $(\beta=0.00987, I_s=0.7741)$ and TB$_2$ $(\beta=0.02944, I_s=0.6707)$,
discussed in the next two Sections.

Without the bilaplacian term there is only one TB point
\cite{2005PhRvL..94f3905G,Damia_PRE}. In that case, the TB point is found only 
asymptotically for the limit $\theta \rightarrow \infty$. As we will discuss 
later, that TB point corresponds in fact to the TB$_1$ point found here. 
Therefore, the  nonlocality has brought this bifurcation to finite parameter 
values allowing us to fully study the different dynamical regimes around the 
TB$_1$ point.

\begin{figure}
\includegraphics[clip]{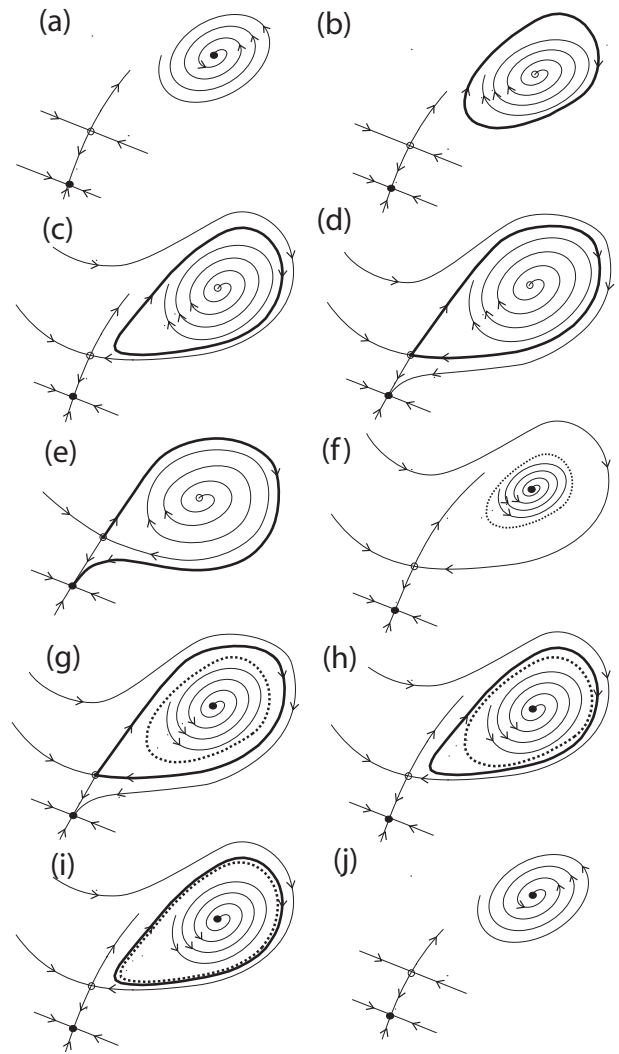}
\caption{Qualitative evolution of the phase space at $I_s \approx 0.8$ and
$\theta=1.23$ for increasing values of $\beta$, corresponding to the different
dynamical behavior in a horizontal cut of Fig.~\ref{Fig::bif_beta_Is}. From (a)
to (j), one goes from stable DS (Region II) to an oscillating DS (Region IV), 
followed by an excitable DS (Region V), a conditional excitable DS (Region VI),
a coexistence of stable DS and oscillatory DS (Region VIII) and 
finally a stable DS (Region III).}
\label{Fig::sketch}
\end{figure}

\subsection{Dynamical scenario around the TB$_1$ point} \label{ssec:tb1}

A Takens-Bogdanov (or double-zero) bifurcation is associated to the presence of
two (non-diagonalizable) degenerate null eigenvalues
\cite{Takens1974,Bogdanov1975}.  Such a bifurcation occurs when, in a line of SN
bifurcations, one of the modes transverse to the center manifold (of the SN
bifurcation) passes through zero, implying that this transverse mode
switches from stable to unstable or vice versa. If this transverse mode is stable
we will denote the SN bifurcation line as SN$^-$, while we use SN$^+$ if this mode
is unstable. Throughout the remainder of this paper, we will use $ ^-$ for bifurcations for which there is a stable emerging solution and $ ^+$ if the emerging solutions are unstable. H$^-$ will describe a supercritical Hopf, and H$^+$ a subcritical one.

Another feature of a TB point is that two new bifurcation lines
emerge from it \cite{Kuznetsovbook}: a Hopf bifurcation line \footnote{The
imaginary part of the eigenvalues at the Hopf bifurcation is singularly zero at
the TB point, in order to have a double zero.} and a saddle-loop (homoclinic)
bifurcation line \footnote{This is an example of the local origin of a global
bifurcation, one of the few situations in which the existence of such a
bifurcation can be established analytically.}. In order to specify whether the cycle that emerges from the saddle-loop bifurcation is stable or unstable, it is useful to define the saddle-quantity $\nu$. For low-dimensional dynamical systems this quantity is given by $\nu=\lambda_s+\lambda_u$, with
$\lambda_u>0$ and $\lambda_s<0$ the  unstable and stable eigenvalues of the
saddle, respectively. The emerging cycle will be stable if  $\nu<0$, and unstable
in the opposite case: $\nu>0$ \cite{Kuznetsovbook}.

At the TB$_1$ point, the saddle-node bifurcation is {\it stable\/} (SN$^-$), the Hopf is supercritical (H$^-$), and the saddle-loop 
creates a stable cycle (SL$^-$). Along the SN$^-$
line a pair of stationary DS are created, one stable (upper branch) and the
other (middle  branch) unstable along a single direction in phase space (thus, a
saddle point in dynamical systems parlance). So, Region II is characterized by
stable DS coexisting with the spatially homogeneous solution.  
A qualitative sketch of the most relevant different kinds of behavior in the system as $\beta$ is
increased  can be found in Fig.~\ref{Fig::sketch} (this corresponds to a
horizontal line in Fig.~\ref{Fig::bif_beta_Is_ZoomTB1} for $I_s\approx 0.8$). In
this figure, panel (a) reflects the behavior inside Region II, where the
DS is the stable focus, the homogeneous solution is the stable node and the
middle branch DS is the saddle.
 
\begin{figure}
\includegraphics[clip]{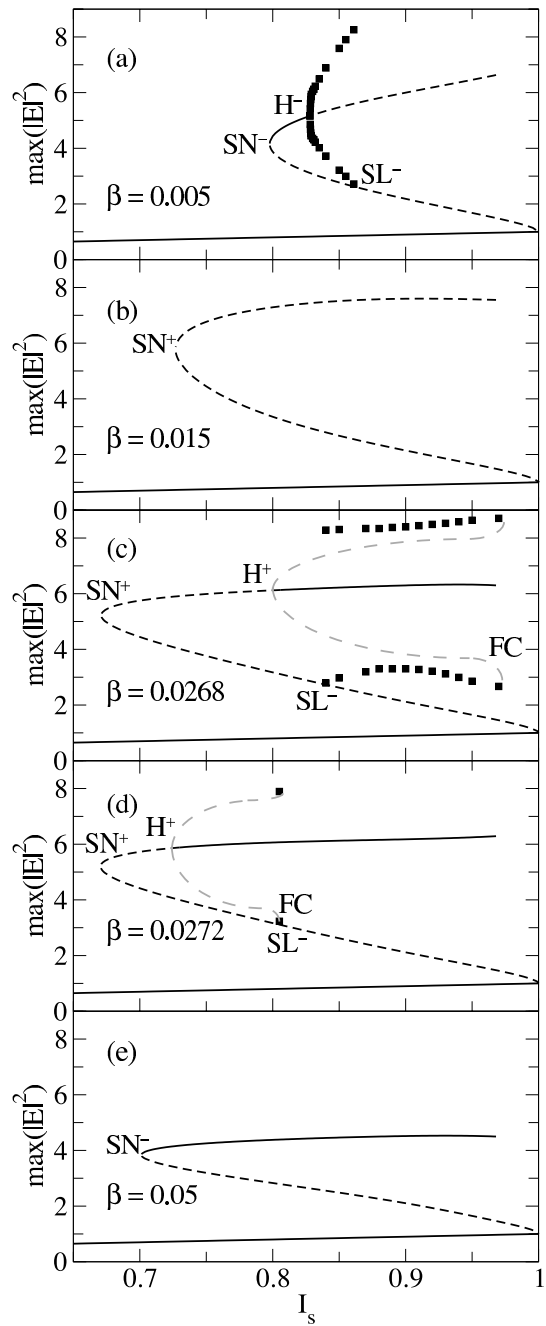}
\caption{Bifurcation diagrams showing the maximum intensity of the DS as
function of the background intensity $I_s$, for: (a) $\beta = 0.005$;
(b) $\beta = 0.015$;  (c) $\beta= 0.0268$; (d) $\beta= 0.0272$; (e) $\beta =
0.05$. The lowest solid line  represents the  stable homogeneous solution. The
lowest dashed line shows the maximum  intensity of the unstable middle branch
DS. Above that line, the upper branch DS is shown, where the solid line
stands for the stable DS and the dashed line for the unstable DS. In (a), (c) and (d) the system exhibits stable oscillatory
behavior (see text). In these cases, the maximum and minimum intensity of the
oscillating DS are depicted as filled squares. The grey dashed line, representing the unstable limit cycle, is only a guiding line for the eye. When the limit cycle touches
the middle branch DS,  a saddle-loop bifurcation occurs. $\theta$ = 1.23.}
\label{Fig::fig_bifurc_branches}
\end{figure}

The upper branch DS solution becomes unstable at the supercritical Hopf
bifurcation, H$^-$, and leads to Region IV, characterized by oscillatory DS,
i.e., autonomous oscillons. Fig.~\ref{Fig::sketch} (b) illustrates the 
behavior past the Hopf bifurcation: the stable oscillatory DS and the unstable
focus in the center can be seen.
Approaching the SL$^-$ line a saddle-loop (homoclinic) bifurcation takes place, in which the
limit cycle (oscillatory DS) becomes a homoclinic orbit of  the saddle (middle
branch DS). The SL$^-$ is a global bifurcation, and cannot be
detected through a local analysis. Thus, in this study it has been determined 
through numerical simulations of Eq.\ (\ref{Eq:LL_beta}). Panel (c) in 
Fig.~\ref{Fig::sketch} illustrates the cycle growing in amplitude and 
approaching the saddle, while in panel (d) the cycle has become
the homoclinic orbit of the saddle.
The approach of the stable cycle to the saddle can 
also be seen quantitatively in panel (a) of Fig.~\ref{Fig::fig_bifurc_branches}: 
in this figure bifurcation diagrams corresponding to vertical cuts in parameter
space, i.e. with $\beta$ fixed (cf. Fig.~\ref{Fig::bif_beta_Is}), are presented. 
Beyond the SL$^-$, the behavior of the system is excitable (Region
V) \cite{2005PhRvL..94f3905G,Damia_PRE}, in particular of type (or class) I 
\cite{IzhikevichIJBC}, as the excitability threshold is the stable manifold 
of the saddle. An excitable excursion is achieved when 
localized perturbations beyond this threshold are applied to the spatially 
homogeneous solution. Fig ~\ref{Fig::sketch} (e) sketches the phase space in
the excitable regime past the saddle-loop bifurcation.

Inside Region V, no stable DS are
found, only unstable DS solutions exist, and the system is excitable for all
values of $I_s$ above the SN$^+$ line. This line which is just the 
continuation of SN$^-$ past the TB$_1$ point was is not observed in 
\cite{2005PhRvL..94f3905G,Damia_PRE} since the TB point without bilaplacian 
term is located at $\theta \rightarrow \infty$. The SN$^+$ line creates a 
saddle-unstable node pair (middle and upper branch, respectively). 
The pair of unstable solutions can be seen in the bifurcation diagram shown in 
Fig.~\ref{Fig::fig_bifurc_branches}(b) corresponding to a vertical cut of the 
parameter plane just to the right of TB$_1$ point.

\subsection{Dynamical scenario around the TB$_2$ point} \label{ssec:tb2}

An even richer scenario is found around the TB$_2$ point. Comparing
Fig.~\ref{Fig::bif_beta_Is_ZoomTB1}(a) and (b), one can see that the TB$_2$ does not
yield the same scenario as around TB$_1$. In TB$_1$ the two lines that
emerge involve stable objects (H$^-$ and SL$^-$ create and destroy,
respectively, a stable limit cycle), while in TB$_2$ the Hopf line is subcritical
(involving an unstable cycle). Furthermore, in the TB$_1$ point the two lines,
H$^-$ and SL$^-$ are tangent to the line SN$^-$ of stable saddle-node
bifurcations while the opposite happens for TB$_2$ (the lines unfolding are
tangent to SN$^+$). 

These differences correspond to a change of sign in a term of the normal form
of the TB bifurcation \cite{Kuznetsovbook}:
\begin{eqnarray}
\begin{array}{rcl} 
\dot{x}&=&y\\
\dot{y}&=&a+b\,x+\,x^2+s\,x\,y\ ,\quad s=\pm 1.
\end{array}
\label{TBnform}
\end{eqnarray}

The case that is most often discussed in the literature leads to the scenario discussed in Section \ref{ssec:tb1}, and corresponds
to $s=-1$. A supercritical Hopf line unfolds from the TB point tangent to the
SN$^-$ segment, and a saddle-loop, in which a stable cycle is destroyed unfolds
tangent to the Hopf line. In the $s=+1$ case, however, a
subcritical Hopf bifurcation line that creates an unstable cycle unfolds tangent
to the SN$^+$ segment, and the unstable cycle is destroyed in a saddle-loop also
tangent to the Hopf bifurcation. This can be seen noticing that changing the sign of $s$ is
equivalent to perform the substitutions $t\mapsto -t$ and $y\mapsto -y$. This
scenario fits nicely with what is observed around the TB$_2$ point, and, thus,
TB$_1$ and TB$_2$ correspond to the two possible cases of Eq. (\ref{TBnform}), 
with $s=-1$ and $s=+1$, respectively. 

Crossing the H$^+$ line coming from Region V, the unstable
(upper branch) DS exhibits a subcritical Hopf bifurcation and becomes a stable focus. The cycle that is created is unstable and has only a
single  unstable direction. In Region VI the system is then bistable (upper
branch DS and homogeneous solutions coexist) and the
unstable cycle is the basin boundary of the upper branch stable DS as
qualitatively illustrated in Fig.~\ref{Fig::sketch} (f). Just after
the bifurcation this basin of attraction is small (since the initial cycle 
amplitude is zero) and grows as one moves away from the bifurcation line. 
This can be seen in the bifurcation diagram displayed in 
Fig.~\ref{Fig::fig_bifurc_branches}(c) corresponding to a vertical cut of 
Figs.~\ref{Fig::bif_beta_Is} and \ref{Fig::bif_beta_Is_ZoomTB1} at $\beta = 0.0268$. The upper branch DS  becomes
stable around $I_s=0.8$. The dashed grey line has been drawn with the
purpose of guiding the eye only. It represents the unstable cycle that we
do not compute. This bifurcation diagram also shows the existence of a stable 
limit cycle (plotted as squares) corresponding to an oscillatory DS. The 
stable limit cycle comes from a fold of cycles (FC) bifurcation, discussed in more 
detail later, which takes place at a larger value of $I_s$. Decreasing $I_s$ the stable limit cycle disappears at a 
saddle-loop bifurcation which takes place when the stable limit cycle becomes
the homoclinic orbit of the saddle (middle branch soliton).
In Fig.~\ref{Fig::fig_bifurc_branches}(c) Region VI corresponds to the values of
$I_s$ limited on the left by the subcritical Hopf (H$^+$) and the right by the saddle 
loop (SL$^+$ and SL$^-$). Precise values for the SL$^-$ obtained from numerical integration of Eq.\ (\ref{Eq:LL_beta}) are plotted as filled squares in 
Fig.~\ref{Fig::bif_beta_Is_ZoomTB1}(b) while the grey SL line joining the points
have been drawn to guide the eye. As we will discuss in the next subsection,
Region VI corresponds to a regime of conditional excitability. 

\begin{figure}
\includegraphics[clip]{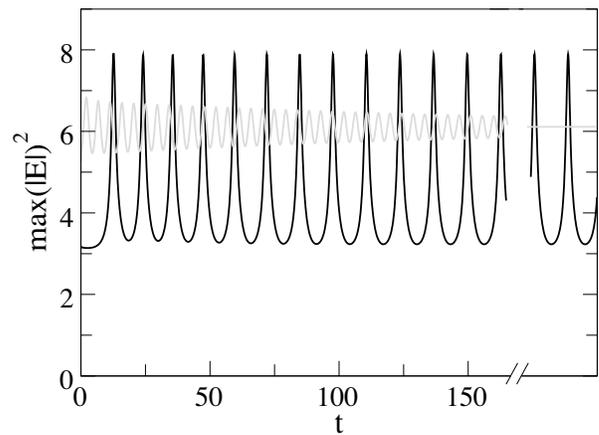}
\caption{Dynamical evolution of the system in the tristable regime, where
the spatially homogeneous fundamental branch (not shown), stable oscillatory 
DS regime (solid line), and stable focus (gray line) coexists.}
\label{Fig::tristability}
\end{figure}

In Fig.~\ref{Fig::fig_bifurc_branches}(c), the FC on the one hand and 
the SL$^-$ on the other limit a new region of {\it tristability} where 
a stationary DS, a oscillatory DS and the homogeneous solution coexist. 
In Fig.~\ref{Fig::bif_beta_Is_ZoomTB1}(b) the tristable region is labeled as VIII.
In Fig.~\ref{Fig::sketch} the panel (g) sketches the phase space at the 
saddle-loop while the panel (h) illustrates the tristable regime. Increasing 
$\beta$ or $I_s$ in parameter space the stable cycle decreases in amplitude 
while the unstable limit cycle increases until both the stable and unstable 
cycles are destroyed in the fold bifurcation (Fig.~\ref{Fig::sketch}(i)).
Figure \ref{Fig::tristability} shows the time evolution in the tristable regime obtained
starting from an initial condition belonging to the basin of attraction of the
limit cycle and from an initial condition within the basin of attraction of the
stable DS. 

\begin{figure}
\includegraphics[clip]{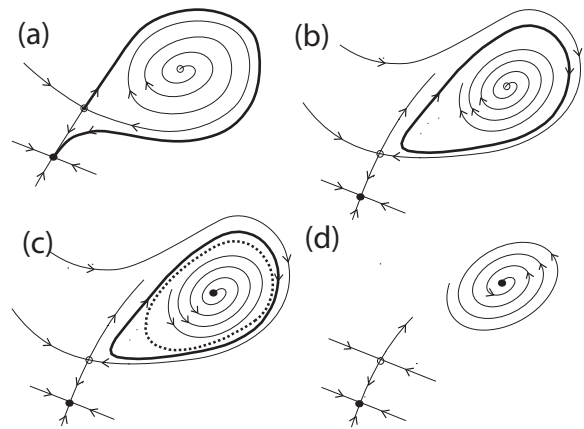}
\caption{A qualitative sketch of the phase space at $I_s \approx 0.9$ and $\theta=1.23$ for increasing values of $\beta$, corresponding to a horizontal cut of Fig.~\ref{Fig::bif_beta_Is_ZoomTB1}(b), is depicted in Fig.~\ref{Fig::sketchbis}. Increasing $\beta$ from (a) to (d), one starts with excitable DS (Region V), going through a region of bistability between a stable limit cycle and the stable homogeneous solution (Region VII), followed by a coexistence of stable DS and oscillatory DS (Region VIII) and 
finally a stable DS (Region III).}
\label{Fig::sketchbis}
\end{figure}

In the phase diagram (Fig.~\ref{Fig::bif_beta_Is_ZoomTB1}(b)) the SL$^-$ line can
also be located to the left of H$^+$ (Region VII). The bifurcation diagram in this case would be similar to the one shown in Fig.\ \ref{Fig::fig_bifurc_branches}(c), except for the fact that the SL$^-$ becomes a SL$^+$ and the H$^+$ takes place at a larger value of $I_s$ than the SL$^-$. Therefore VII is a region of
bistability where a stable limit cycle corresponding to the oscillatory upper 
branch DS coexist with the homogeneous solution while the steady state upper
branch DS is an unstable focus. A qualitative sketch of the phase space when going from region V through VII and VIII is depicted in Fig.~\ref{Fig::sketchbis}.

The last region in the phase diagram to be described is Region III, located to
the right of the SL$^+$ and FC lines. The phase space corresponding to this
broad region is sketched in Figs.~\ref{Fig::sketch}(j) and \ref{Fig::sketchbis}(d). The upper 
branch stationary DS is a stable point and coexists with the homogeneous 
solution. Fig.~\ref{Fig::fig_bifurc_branches}(e) shows a quantitative
bifurcation diagram corresponding to a vertical cut of Fig.~\ref{Fig::bif_beta_Is} for $\beta=0.05$, to the right of the 
TB$_2$. For $I_s<0.7$
only the homogeneous solution exists (Region I) while above $I_s=0.7$ one enters
in Region III of coexistence of the stable DS with the homogeneous solution.

We now analyze in detail the fold of limit cycles bifurcation. It appears 
at a secondary codimension-2 point named
resonant side-switching, where the fold and the saddle-loop bifurcation, that
delimit the region of tristability, coalesce \cite{Chow90,ChampneysKuznetsov}. 
This occurs, roughly, around $\beta= 0.0272$ and $I_s=0.82$. 
Fig.~\ref{Fig::fig_bifurc_branches}(d) depicts a bifurcation diagram close to 
this codimension-2 point, for which the region of existence of stable limit 
cycles is quite narrow: the represented square is very close both to the 
saddle-loop bifurcation and to the fold of cycles. In the phase diagram shown in
Fig.~\ref{Fig::bif_beta_Is_ZoomTB1}(b) this codimension-2 corresponds to the
point where the FC and the SL (SL$^-$ and SL$^+$) lines meet.

\begin{figure}
\includegraphics[clip]{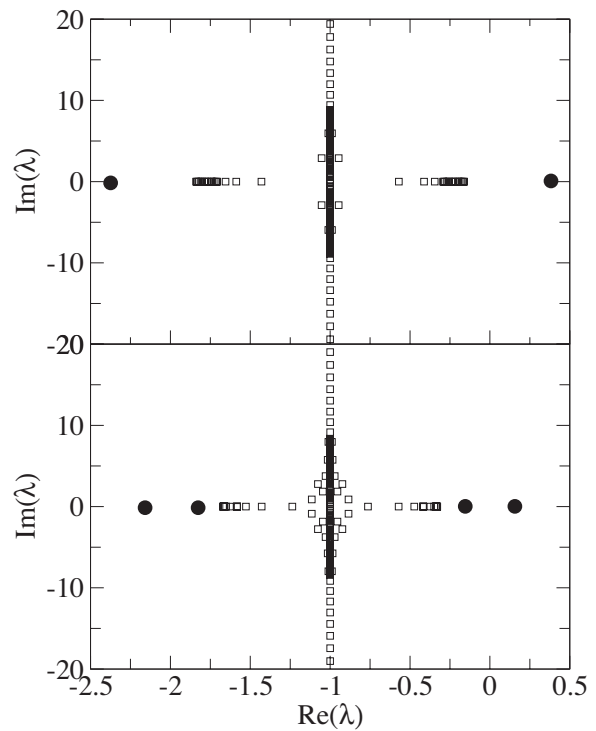}
\caption{The spectrum of the middle-branch DS at the
saddle-loop  bifurcation for (a) $\beta = 0.0268$ and $I_s = 0.84$, and (b)
$\beta = 0.0294$  and $I_s = 0.671$. The filled black circles are the eigenvalues
corresponding to the modes with a localized spatial profile. $\theta = 1.23$.}
\label{Fig::Saddle_index_Homoclinic}
\end{figure}

The occurrence of the resonant side-switching bifurcation is related to the
eigenvalue spectrum of the saddle, namely, the saddle quantity of the middle
branch DS approaches zero, and the cycle emerging from the saddle-loop changes from
''stable'' to ''unstable''. Close to TB$_2$ point the
saddle-loop bifurcation must destroy an unstable cycle (transition from Region
VI to III), hence $\nu>0$, while
after the fold has taken place it destroys a stable cycle (transition from
Region VIII to VI), so $\nu<0$.
Fig.~\ref{Fig::Saddle_index_Homoclinic} shows the spectrum of the middle branch DS for two
different sets of parameters, one after the formation of the fold (saddle-loop
bifurcation with a stable cycle, top panel) and one very close to the TB$_2$
point (saddle-loop bifurcation with an unstable cycle, bottom panel). The
eigenvalues of the modes relevant for the dynamics are highlighted with filled
circles, and indeed, the saddle quantity is positive $\nu>0$, albeit quite 
small ($\nu\sim 0.003$) in the bottom panel, compatible with the fact that it 
involves an unstable cycle. One can observe that close to the SL$^+$ that emerges from the TB$_2$ point, where the saddle quantity is positive, four localized modes play a role in the dynamical behavior of the system. However, when moving away from the SL$^+$ line, only two localized modes determining the dynamics remain. So, in conclusion, we can say that the system is essentially two-dimensional for the same reasons as discussed in Ref.\ \cite{Damia_PRE}, except close to the  SL$^+$ line, originating from the TB$_2$ point.

\subsection{Excitability and conditional excitability}

When crossing a saddle-loop bifurcation from a region where a stable oscillatory
DS exist one enters in a excitable regime. As discussed in Subsection~\ref{ssec:tb1}, this scenario takes place when going from Region IV (where the
oscillatory DS is stable) to Region V where DS exhibit excitability. 
Of course it is also possible to enter in Region V from the other side, namely 
from the side of the TB$_2$ point. In fact a similar scenario is found when
going from Region VII where the oscillatory DS is stable to Region V. In any
case the stable manifold of the saddle (middle branch DS) plays the role of
excitability threshold, so the excitable response is triggered only by localized
perturbations of the homogeneous solution that bring the system beyond this
threshold. Excitability is of Class I \cite{IzhikevichIJBC}, 
characterized by long response times for perturbations that leave the 
trajectory close to the saddle in phase space.

Close to the TB$_2$ point there is another region, VI, where one can enter
crossing a saddle-loop line. However, the dynamical behavior in Region VI is 
qualitatively different from Region V since the upper branch LS is an 
unstable focus in the last one, while it is stable in the former. As discussed 
before, the upper branch LS has
been made stable by the subcritical Hopf bifurcation H$^+$ that separates 
Region V from VI. The H$^+$ bifurcation generates also an unstable limit cycle,
which is not present in Region V.
Therefore, although the transition from Regions III or VIII to Region VI goes
through a saddle-loop bifurcation, the scenario must be qualitatively different
from the one discussed above which leads to the {\it usual\/} excitability found
in Region V. In Region VI, one finds a regime of conditional excitability, in 
which the DS is simultaneously excitable and bistable. The excitable behavior 
is also Class I.

\begin{figure}
\includegraphics[clip]{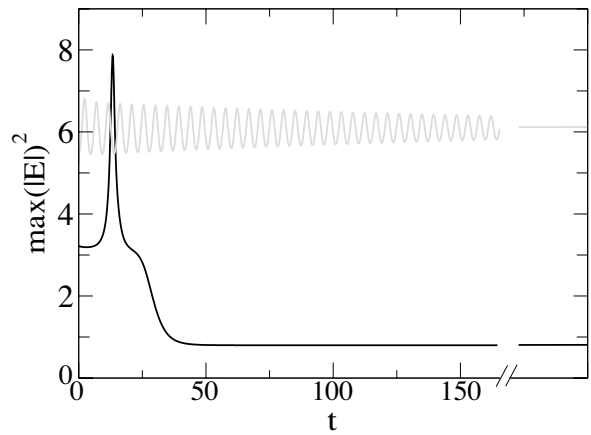}
\caption{The conditional excitability regime is illustrated for $\beta = 0.0272$
and $I_s=0.8$. The solid line (excitable trajectory) corresponds to time
evolution of the maxima of the peak after a localized perturbation of the
homogeneous state that brings the system beyond the excitability threshold but 
outside the basin of attraction of the upper branch DS. The grey line
corresponds to a stronger localized perturbation that brings the system inside 
the boundary of attraction of the upper branch DS.}
\label{Fig::conditexcit}
\end{figure}

To clarify what conditional excitability means, we refer to the phase space
sketched in Fig.~\ref{Fig::sketch}(f). In this situation, while as usual,
perturbations of the homogeneous solution that are not able to cross the 
excitability threshold (stable manifold of the saddle) 
lead to normal relaxation, there are two possible different dynamical responses for
supra-threshold perturbations. If a localized perturbation of the homogeneous 
state brings the system inside the basin of attraction of the stable focus, 
namely inside the unstable cycle, 
the system jumps from the fundamental solution to this attractor. Therefore
after this perturbation the system relaxes to the stable DS in a oscillatory 
way. The grey line in Fig.~\ref{Fig::conditexcit} shows the dynamical 
evolution of the maxima of the peak in this situation.
Instead, for localized perturbations of the homogeneous solution which bring the
system beyond the stable manifold of the saddle but outside the unstable
cycle, the response is excitable. The system exhibits a large response 
corresponding to a circulation around the unstable limit cycle before 
returning to the stable homogeneous solution. The black line in 
Fig.~\ref{Fig::conditexcit} shows the time evolution of maxima of the peak for a
excitable trajectory.

So, in summary, the dynamical response of perturbations is more complex 
than simply sub- and supra-threshold, and for the latter type of perturbations 
two possible regimes are possible.
When going from Region VI to Region V at the H$^+$ line, the upper branch DS 
becomes unstable and the unstable limit cycle responsible for the conditional 
excitable response to supra-threshold perturbations disappears, so the 
conditional excitability becomes a usual one.

\section{Concluding Remarks} 

\begin{figure}
\includegraphics[clip]{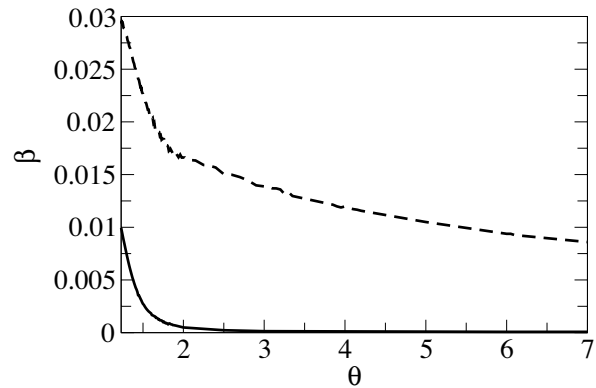}
\caption{Location ($\theta_{TB},$$\beta_{TB}$) of the two TB points as the detuning is increased. The solid line and the dashed line correspond to the TB$_1$ and TB$_2$ point, respectively. }
\label{Fig::fig_evol_TB}
\end{figure}

We have studied the nonlinear dynamical behavior of 2D localized structures in a
model for an optical cavity filled by a Kerr nonlinear medium and a left-handed
metamaterial \cite{Gelens_PRA_2007}. The model is a generalization of the 
Lugiato-Lefever equation \cite{LL-1987}, and includes higher order spatial 
effects arising from the weakly nonlocal response of the metamaterial. In
this system, we have shown the existence of regions with stationary,
oscillating and excitable localized structures. Furthermore, we have shown that
the different bifurcation lines originate from two Takens-Bogdanov (TB)
codimension-2 points, which is a strong signature for the presence of a
homoclinic bifurcation. This homoclinic bifurcation offers a route to excitable
behavior of the 2D localized structures. Finally an extra secondary
codimension-2 point (resonant side-switching bifurcation) creates a fold of
cycles that leads to two new regimes, one of tristability and one of conditional
excitability.

Without the nonlocal terms the different dynamical regimes of the DS were 
organized by a TB point located at the limit of infinite detuning, where the 
Lugiato-Lefever equation reduces to the (conservative) Nonlinear Schr\"odinger 
Equation (NLSE) \cite{2005PhRvL..94f3905G, Damia_PRE}. Here, we demonstrate that the presence of higher order spatial effects brings two TB points to finite parameter values. In Fig.\ \ref{Fig::fig_evol_TB}, we provide further evidence that TB$_1$ is unfolding from the NLSE at the limit $\beta \rightarrow 0$ and then $\theta \rightarrow \infty$ of Eq.\ (\ref{Eq:LL_beta}). This hypothesis is supported by the fact that 2D solitons in the NLSE have at least a twofold degeneracy \cite{Skryabin_2002}. Furthermore, Fig.\ \ref{Fig::fig_evol_TB} also gives some evidence that TB$_2$ comes from a certain conservative limit at $\theta \rightarrow \infty$, $\beta \rightarrow 0$ of Eq.\ (\ref{Eq:LL_beta}). However, if this is a singular limit, this does not necessarily imply that TB$_1$ and TB$_2$ have to evolve towards the same point in the conservative limit. The numerical evidence given in Fig.\ \ref{Fig::fig_evol_TB} is of course not fully conclusive, and this unfolding of both of the TB points is presently under investigation.

\begin{acknowledgments}
We thank D. Paz\'o for useful discussions. This work was supported by the
Belgian Science Policy Office under grant No.\ IAP-VI10, by the Spanish Ministry
of Education (MEC) and FEDER under grants No.\ FIS2004-00953 (CONOCE2),
TEC2006-10009 (PhoDeCC), FIS2006-09966 (SICOFIB), and FIS2007-60327 (FISICOS), and by the Govern Balear
under grant No. PROGECIB-5A. LG is a PhD Fellow and GV is a Postdoctoral Fellow
of the Research Foundation - Flanders (\textsc{FWO} - Vlaanderen).
\end{acknowledgments}


\end{document}